# Building a Controlled Vocabulary for Standardizing Precision Medicine Terms

Meng Wu
Institute of Medical Information & Library, Chinese Academy of Medical Sciences
Beijing, 100020
China
wu.meng@imicams.ac.cn

Yan Liu
Institute of Medical Information & Library, Chinese Academy of Medical Sciences
Beijing, 100020
China
liu.yan@imicams.ac.cn

Hongyu Kang
Institute of Medical Information & Library, Chinese Academy of Medical Sciences
Beijing, 100020
China
kang.hongyu@imicams.ac.cn

Si Zheng
Institute of Medical Information & Library, Chinese Academy of Medical Sciences
Beijing, 100020
China
zheng.si@imicams.ac.cn

Jiao Li
Institute of Medical Information & Library, Chinese Academy of Medical Sciences
Beijing, 100020
China
li.jiao@imicams.ac.cn

Li Hou*
Institute of Medical Information & Library, Chinese Academy of Medical Sciences
Beijing, 100020
China
hou.li@imicams.ac.cn

## ABSTRACT

Rapid advances of technology and development of research in precision medicine domain have led to the production of different types of biomedical data. Standard medical vocabularies were shown to be limited in dealing with such heterogeneous data and consequently, new controlled vocabulary for data integration and normalization has been proposed. In this study, the precision medicine vocabulary (PMV), which is a controlled vocabulary for terms used in precision medicine, is built based on the method of data integration in Unified Medical Language System (UMLS). It now covers ten top semantic types of disease, drug, gene, gene variation and so on. In total of 1,372,967 concepts and 4,567,208 terms have been integrated from widely used databases related with precision medicine.

## CCS CONCEPTS

• **Information systems** → **Information integration;** • **Information systems**→Data management systems;

## KEYWORDS

controlled vocabulary, data integration, data standard.

## 1 INTRODUCTION

A controlled vocabulary plays a fundamental role in data mining and knowledge discovery. In precision medicine studies, multidisciplinary scientists have made efforts towards discovering disease mechanism and providing diagnosis and treatment evidences for precise clinical practices. The scientists may use diverse terms to describe their research findings. And the forms of terms used in different biomedical databases can be described in many ways. So, it's essential and necessary to build a controlled vocabulary for standardization and integration of precision medicine terms. And the main goal of data integration is to extract additional biological knowledge from multiple datasets that cannot be gained from any single dataset alone [1].

The existing medical comprehensive resources, such as UMLS [2], has a large volume of medical concepts as we all know. But the classification system of the concepts is not detailed enough, and there is a lack of integration of large data at the level of disease molecular mechanism, which are the important aspect considered in precision medicine.

The Precision Medicine Vocabulary (PMV) has been developed as a standard thesaurus for integrating and representing the data in human precision medicine domain with consistent, reusable and sustainable descriptions of human disease, genomic molecular, phenotype characteristics and other related medical vocabulary through collaborative efforts from multidisciplinary researchers. The building process also indicates the scalability and flexibility of PMV. And the PMV provides a standard and structural support for the Precision Medicine Knowledge Base (PMKB) built based on it.

## 2 METHODS

### 2.1 Scope of Precision Medicine Definition

PMV is an open source controlled vocabulary of precision medicine. The concept of precision medicine was first given profile by a publication from the National Research Council, which says a new data network that integrates emerging research on the molecular makeup of diseases with clinical data on individual patients could drive the development of a more accurate classification of diseases and ultimately enhance diagnosis and treatment [3]. And the precision medicine research initiative in American, which is called ALL of US research program now, focuses on the intersection of environment, lifestyle, and biology. To give a better description of data in precision medicine domain, the data of exposome, signs and symptoms, genome and other data related with disease molecular mechanism and etiology are mainly considered by us.

And the scope is also extended and defined to meet the needs of the whole project of precision medicine knowledge base. The Precision Medicine Knowledge Base, which is the National Key Research and Development Program of China, aims to construct a reliable knowledge base of precision medicine for massive data analysis and integration. The identification and integration of gene, disease, drug, mutation, phenotypic abnormality, pathway, and so on, supported by the PMV, are also the key scope of the data in the PM knowledge base.

To provide better services and management of vocabulary, the terms in PMV are organized into ten top data semantic types, these are anatomical structure, gene, gene product, mutation, cell, disease, phenotype, biomedical pathway, biological function, chemical and drug with uniform identification. The PMV project continues to improve the representation of all data in human precision medicine domain with the addition of new PM terms with the needs of curation, term requests and collaborative.

## 2.2 Controlled Vocabularies Collection

Taking the PM scope into fully consideration, we selected the data in related databases as the foundational vocabulary based on UMLS. Firstly, we extracted the concepts from UMLS according to whether or not their semantic types are within the PM scope, for example, the data in Anatomical Structure, Chemical, Clinical Drug, and Sign or Symptom were all acquired by us. Secondly, the terms which belong to non-human species and non-English languages were removed to meet the current needs of the knowledge base. Thirdly, we limited the number of the sources of concept to 52 vocabularies after removing the vocabularies whose amounts of terms are very few or subjects are not related with our scope.

In the end, the foundational vocabulary contains both comprehensive vocabularies such as Medical Subject Headings (MeSH), National Cancer Institute Thesaurus (NCIt) and Systematized Nomenclature of Medicine Clinical Terms (SNOMED CT), and databases in specific domain such as HUGO Gene Nomenclature Committee (HGNC) and Online Mendelian Inheritance in Man (OMIM) for gene, Human Phenotype Ontology (HPO) for human phenotype, DrugBank and RxNorm for drug. The selected data were reorganized based on the organization format described in the later section.

## 2.3 Heterogeneous Biomedical Resources Integration

For giving better representation of the PM domain, we integrated some biomedical resources such as DrugBank, ClinVar and NCBI Gene with the foundational vocabulary by utilizing a series of mapping and integration strategies. DrugBank has been included in the vocabulary already as mentioned above, but it is found that UMLS integrates it incompletely. Therefore, further integration of DrugBank is necessary. The detailed data match and integration workflow is portrayed in figure 1.

These resources are arranged to be integrated into the vocabulary one after another. As for one resource, first, the name and synonyms of each concept in the resource are parsed and transformed into concept-term form according to the content and structure of the resource. For one gene in NCBI gene database, the symbol of the gene is treated as the concept and the synonyms and other full names are treated as the terms.

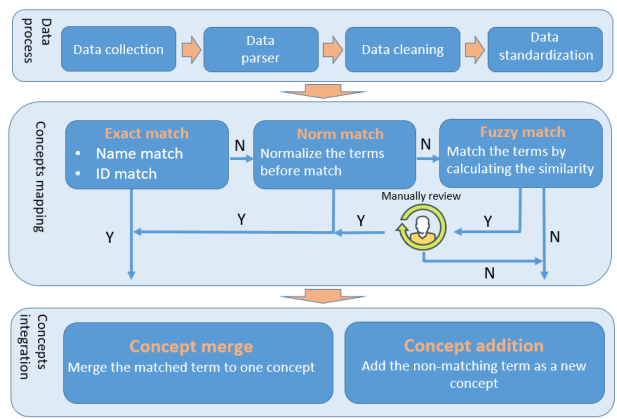

**Figure 1: The Detailed Integration Workflow of the Heterogeneous Biomedical Resources. In which 'Y' and 'N' respectively represent the positive and negative results of the judgement.**

In order to match the terms as many as possible, three kinds of match strategies are performed in turn between each term of the resource and each term in the latest version of PMV. The exact match utilizes the name, ID of the term and the name of the resource to supply an accurate mapping. The norm tool [4] which is one of the Lexical Tools of UMLS is used in the norm match process. We use the tool to generate the normalized string for the terms before matching them. The normalization process involves stripping possessives, replacing punctuation with spaces, removing stop words, lower-casing each word, breaking a string into its constituent words, and sorting the words in alphabetic order. After the fuzzy match, there is a manually review process. The experts of biomedical domain will check the top five matched results of fuzzy match, and pick a most correct match or fail them all. The matched term and other terms in the same concept will be merged into the concept existing in PMV. The concept in which all the terms are unmatched will be added into PMV as a new concept.

After executing those match processes, 4,571 drug concepts and 150,629 terms in DrugBank which had not previously been identified and integrated in UMLS were added into PMV. For example, 5 additional DrugBank terms were identified and integrated into PMV for the drug 'Herceptin', which is a therapeutic drug for patients with breast or metastatic stomach cancer whose tumors overexpress the HER2 gene [5]. As for NCBI Gene and ClinVar, which are the new resources to UMLS, were added into PMV through mapping with existing genes and mutations mainly in OMIM and HUGO in PMV. It provides an increase of 21,172 concepts and 220,328 terms in gene and 294,712 concepts and 316,630 terms in mutation.

## 2.4 Vocabularies Organization

These vocabularies are organized in a standard format referring to the term organization method of UMLS Metathesaurus. As we know, UMLS utilizes many data files which are organized in RRF and ORF formats to manage the biomedical and health related concepts, their various names, and the relationships among them.

As for term identification, we follow and simplify the assignment of several types of unique identifiers used in UMLS to reorganize the concepts and terms. In PMV, we use MCID for concept identification, use MAID for term identification, and use MTID for class identification. Each concept is linked to at least one MAID (term), and only one term among them is chosen as the preferred term for this concept according to the priority of the source vocabulary of the term which is given by UMLS and extended by US. The digits of each type of code are designed based on the actual demand of the knowledge base. The vocabulary coding formats are listed in table 1.

**Table 1: Vocabulary Coding Formats**

| Name | Abbreviation | Format | Example |
|---|---|---|---|
| Concept ID | MCID | MC+8 digits | MC00001175 |
| Term ID | MAID | MA+8 digits | MA00019781 |
| Type ID | MTID | MT+8 digits | MT00000035 |

The metadata of the vocabulary also include the source abbreviation and the code in source. TTY, which is the term type in source description in UMLS, is reused by matching the terms to appropriate term types and extended by defining new term types for them to record the feature of the original data.

Except for the main concept table, there are tables describing the source and version of each vocabulary and semantic type of each concept in PMV. The semantic type will be discussed in the later section.

## 2.5 Semantic Types Reorganization

As the scope of PMV is huge and diverse, we use semantic types to provide classification and semantization for the concepts in PMV.

According to the ten top semantic types we considered above and the semantic types in UMLS, some classification systems and terms in widely used biomedical resources were reused. We selected and obtained the terms and hierarchies from data resources including MeSH, NCIt, HPO, ClinVar, and NCBI gene. These terms and hierarchies are integrated together into a hierarchy structure to organize the vocabulary. The structure of the top level in this hierarchy is indicated in figure 2.

At the top hierarchy, the linkage between concepts and sematic types were inherited from UMLS. For the lower and detail hierarchies, a mapping process between the type and the terms was conducted to link the concepts and sematic types as many as possible. And the mutation type in ClinVar and gene type in NCBI gene are reused as the standard classifications directly in PMV to provide an effective universal data classification.

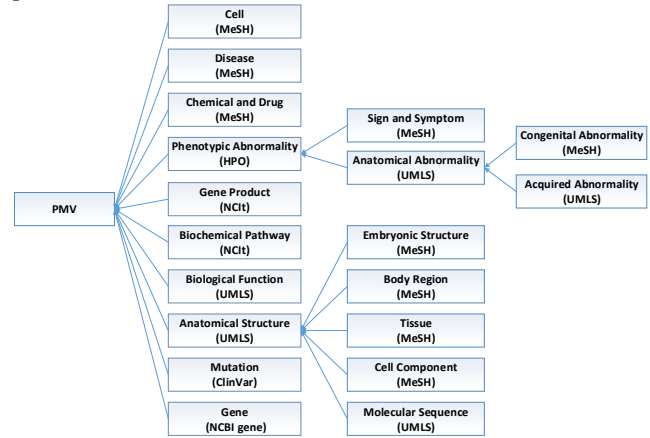

**Figure 2:** The top level hierarchical structure of semantic types in PMV. (Terms and hierarchies imported from other biomedical resources are indicated by the abbreviations inside the parentheses. All the arrows indicate the "Subclass_of" relationship.)

## 3 RESULTS AND DISCUSSION

The latest version of PMV contains 1,372,967 concepts and 4,567,208 terms which covers ten top semantic types in precision medicine domain. To assess PMV in an effective and comprehensive way, we compare the concepts in PMV and UMLS in both dimensions of quantity and scope (Table 2).

**Table 2: The numbers of the concepts in PMV and UMLS under ten top semantic types.**

| TOP CLASS | PMV | UMLS |
|---|---|---|
| Anatomical Structure | 97,653 | 197,162 |
| Phenotypic Abnormality | 33,929 | 13,854 |
| Biochemical Pathway | 2,157 | - |
| Cell | 5,462 | 5,570 |
| Biologic Function | 94,711 | 246,575 |
| Chemical and Drug | 876,306 | 975,604 |
| Disease | 137,419 | 141,314 |
| Gene | 66,022 | 46,948 |
| Mutation | 320,672 | 25,715 |
| Gene Product | 104,124 | 143,375 |

By means of comparing the statistics in ten different dimensions, a number of observations apparent. UMLS has a comprehensive vocabulary which contains terms of multiple species and multiple languages, therefore, it's not surprise that UMLS has more concepts in most semantic types compared with PMV. The PMV concepts in gene, mutation, and pathway are more than UMLS concepts, which indicates that PMV may perform better in precision medicine domain. We defined the scope of the phenotypic abnormality in PMV and reorganized the concepts in it, but it is not appropriate for UMLS and the scope of phenotypic abnormality in UMLS is indistinct. So phenotypic abnormality concepts in PMV are more than that in UMLS.

Alternatively, PMV has good coverage of concepts in biomedicine domain such as gene, mutation and pathway, but has only moderate coverage of concepts in basic medicine domain. This indicated areas of strength and improvement for PMV to cover additional concepts.

## 4 CONCLUSIONS

In summary, we built a controlled vocabulary for integrating and standardizing the terms used in precision medicine research. The large number of source vocabularies included in PMV indicates the role of PMV in interoperability in the PM domain. And the vocabulary provides good scalability and flexibility for users.

In the future, the classification of terms in PMV should be perfected in a more detailed and effective way according to the needs of clinic treatments and scientific research. Secondly, new vocabularies in biomedical domain will be considered to be added into PMV, for example Pathway Common in pathway domain. And more properties of the terms and the semantic relationships between terms will be annotated and added into PMV. Thirdly, an automated vocabulary integration tool should be developed to provide a convenient and effective way for term mapping, term merging, term exporting and version management.

Now the data are stored in MySQL databases for data management. The data of main concepts, the semantic types, and other related information are all organized in table formats. The PMV will be perfected continuously and opened in various data formats soon.

## ACKNOWLEDGMENTS

The authors would like to thank Junlian Li, Xiaoying Li, Yujing Ji, Panpan Deng and Haixia Sun for their contribution to this work. This work is supported by the National Population and Health Scientific Data Sharing Program of China, and by the National Key Research and Development Program of China (Grant No. 2016YFC0901901).